\documentclass[a4paper,11pt]{article}
\usepackage{pos}

\title{Learning Trivializing Flows}
\author*[a]{David Albandea}
\author[b]{Luigi Del Debbio}
\author[a]{Pilar Hernández}
\author[b]{Richard Kenway}
\author[b]{Joe Marsh Rossney}
\author[a]{Alberto Ramos}

\affiliation[a]{Instituto de Física Corpuscular (CSIC -- University of
Valencia), Parque Científico, C/Catedrático José Beltrán, 2, 46980, Paterna,
Valencia, Spain}
\affiliation[b]{Higgs Centre for Theoretical Physica, School of Physics and
Astronomy, The University of Edinburgh, Edinburgh EH9 3FD, UK}

\emailAdd{david.albandea@ific.uv.es}

\abstract{The recent introduction of machine learning techniques, especially
normalizing flows, for the sampling of lattice gauge theories has shed some hope
on improving the sampling efficiency of the traditional HMC algorithm.  Naive
use of normalizing flows has been shown to lead to bad scaling with the volume.
In this talk we propose using local normalizing flows at a scale given by the
correlation length. Even if naively these transformations have a small
acceptance, when combined with the HMC algorithm lead to algorithms with high
acceptance, and also with reduced autocorrelation times compared with HMC.
Several scaling tests are performed in the $\phi^{4}$ theory in 2D.}

\FullConference{%
  The 39th International Symposium on Lattice Field Theory (Lattice2022),\\
  8-13 August, 2022 \\
  Bonn, Germany \\
  Report number: IFIC/22-36
}

\begin{document}
\maketitle

\begin{figure}[ht]
	\centering
    \includegraphics[width=\linewidth]{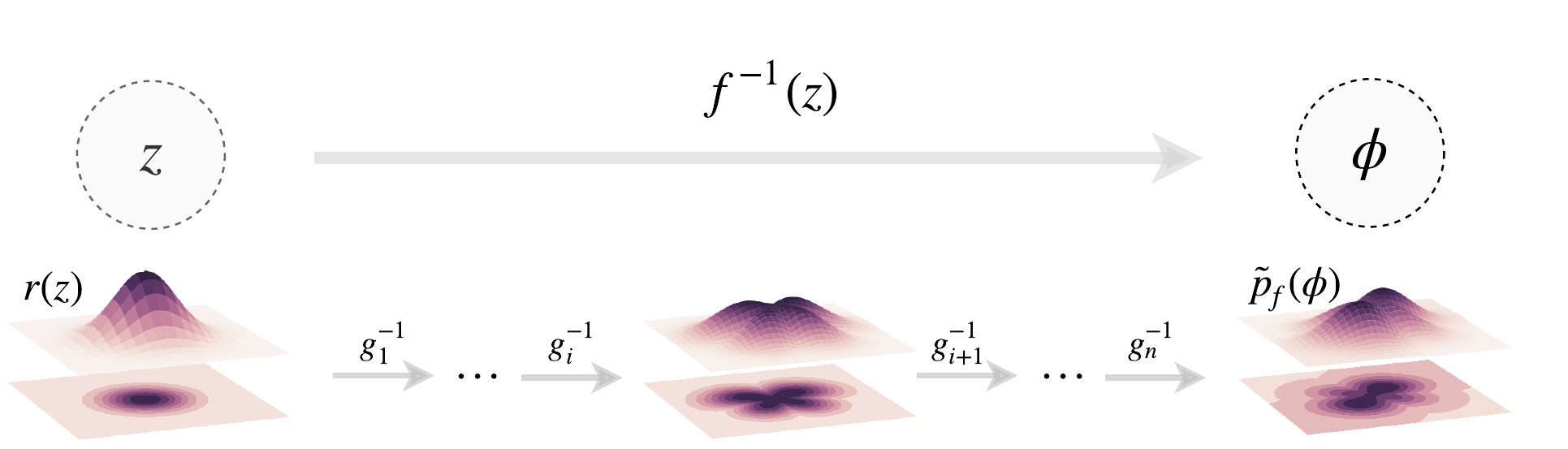}
    \caption{Extracted from \cite{Albergo2019}: normalizing flow sketch. A set
        of configurations $z_{i}$ is generated from a trivial probability
        distribution $r(z)$; a neural network model $f$ is used to generate a
        new set of configurations via $\phi = f^{-1}(z)$, which follow
        $\tilde{p}_{f}(\phi) \approx p(\phi)$, with $\tilde{p}_{f}$ the model's
        output distribution and $p$ the target distribution.}
    \label{fig:nom-flows}
\end{figure}

\section{Introduction}\label{sec:introduction}

\subsection{Normalizing flows}

Normalizing flows are a machine learning sampling technique introduced to lattice
field theories in \cite{Albergo2019}. As sketched in Fig.\ref{fig:nom-flows},
they propose a flow-based Markov Chain Monte Carlo algorithm whose workflow is
\begin{enumerate}
    \item
        Generate a set of configurations $z_{i}$ following a trivial probability
        distribution $r(z)$.
    \item
        Apply a function $f^{-1}$ to all configurations $z_{i}$, obtaining a new
        set of configurations $\phi_{i}$ via $\phi = f^{-1}(z)$, following a new
        probability distribution $\tilde{p}_{f}(\phi)$. $f$ is a neural network
        which has been trained so that the new probability distribution
        $\tilde{p}_{f}$ is a similar as possible to our target distribution $p
        \propto e^{-S}$, the distribution of the theory to be studied.
    \item
        Use a Metropolis--Hastings accept-reject to correct for the bias in the
        approximation $\tilde{p}_{f} \approx p$.
\end{enumerate}
The training of the neural network $f$ is done by minimizing the
Kullbach-Leibler (KL) divergence,
\begin{align}
    D_{\text{KL}}(\tilde{p}_{f} || p) = \int \mathcal{D}\phi \;
    \tilde{p}_{f}(\phi) \log \frac{\tilde{p}_{f}(\phi)}{p(\phi)},
\end{align}
which is a statistical distance satisfying $D_{\text{KL}}(\tilde{p}_{f} || p)
\ge 0$ and $D_{\text{KL}}(\tilde{p}_{f} || p) = 0 \iff \tilde{p}_{f}
= p$. Therefore, after training the network $f$ is expected to be an approximate
trivializing map \cite{LuscherTrivializingMaps}.

\begin{figure}[!t]
	\centering
    \includegraphics[width=0.43\linewidth]{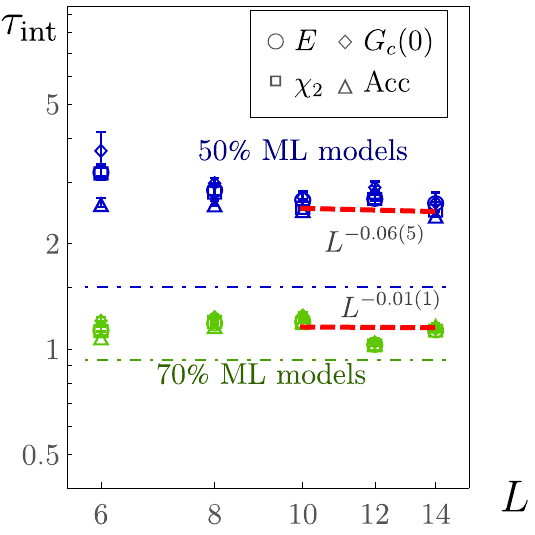}
    \includegraphics[width=0.49\linewidth]{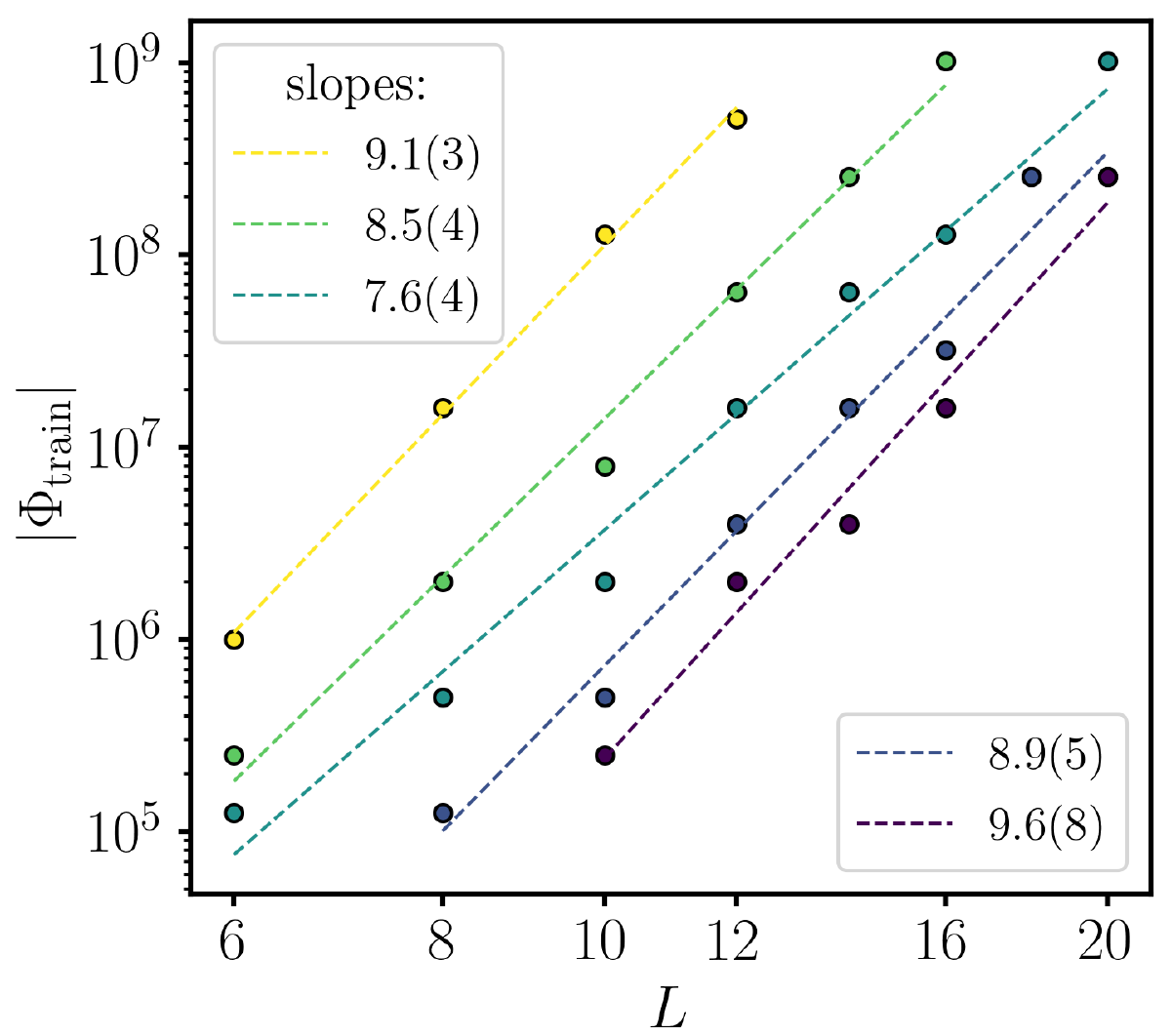}
    \caption{(Left) Extracted from \cite{Albergo2019}: autocorrelation time
        scaling for different observables using normalizing flows with networks
        trained to have Metropolis acceptances of 50\% and 70\%.  (Right)
        Extracted from \cite{DelDebbio2021}:  scaling of the number of
        configurations needed to train a network up to an acceptance of 70\%.}
    \label{fig:scaling}
\end{figure}

As shown in Fig.\ref{fig:scaling} (left), the good thing about this algorithm is
that autocorrelation times in the resulting Markov chain do not scale when
taking the continuum limit if the neural networks are trained up to the same
reference acceptance, meaning that the cost of producing configurations does not
scale; on the other hand, autocorrelations with the Hybrid Monte Carlo (HMC)
algorithm are expected to scale as $\sim \xi^2$. 

However, if one wants to study the scaling of the total computational
cost, one needs to analyze the training costs as well, since
\begin{align}
    \text{total cost} = \text{configuration production cost}  + \text{network
    training cost}.
\end{align}
It was shown in \cite{DelDebbio2021} that the cost of keeping a reference
Metropolis acceptance of 70\% seems to scale approximately as $\sim \xi^{8}$ (see
Fig.\ref{fig:scaling}), indicating a transfer of the critical slowing down
problem from the production of configurations to the training cost of the
networks.

In what follows we want to address whether one can still benefit from
normalizing flows keeping the training costs low.

\section{Learning Trivializing Flows}

\subsection{The algorithm}

The main idea is to use Lüscher's trivializing flows
\cite{LuscherTrivializingMaps}, so that we can use the normalizing flows to
\emph{help} the HMC algorithm, rather than replacing it. For example, let's
consider the partition function of our target theory,
\begin{align}
    Z = \int \mathcal{D}\phi\; e^{-S[\phi]}.
\end{align}
We can make a change of variables $\phi \to \tilde{\phi}$ using our trained
network $f$,
\begin{align}
\tilde{\phi} = f(\phi), \quad \phi  = f^{-1}(\tilde{\phi}), \quad \mathcal{D}\phi  = \left| \frac{\partial f^{-1}(\tilde{\phi })}{\partial
\tilde{\phi}}\right| \mathcal{D}\tilde{\phi } \equiv \det J\left[f^{-1}(\tilde{\phi
})\right] \mathcal{D} \tilde{\phi},
\label{eq:variable-trafo}
\end{align}
so that the partition function becomes
\begin{align}
Z = \int_{ }^{ } \mathcal{D} \tilde{\phi }\; e^{-S[f^{-1}(\tilde{\phi })] + \log \det J[f^{-1}(
\tilde{\phi } )]} \equiv \int_{ }^{ } \mathcal{D}\tilde{\phi} \;
e^{-\tilde{S}[\tilde{\phi}]},
\end{align}
where we have defined the new action
\begin{align}
    \tilde{S}[ \tilde{\phi} ] \equiv S[f^{-1}(\tilde{\phi })] - \log \det J[f^{-1}(
\tilde{\phi } )].
\end{align}
This new action is a combination of the old action $S[ \phi ]$ and the logarithm
of the Jacobian of the variable transformation of Eq.\eqref{eq:variable-trafo}.
If the Jacobian cancels out part of the action, then the probability
distribution $e^{-\tilde{S}[ \tilde{\phi} ]}$ might be easier to sample from
than $e^{-S[ \phi ]}$, and using HMC with the new action might yield lower
autocorrelation times. The workflow of the algorithm would then be
\begin{enumerate}
    \item
        Train the network $f$ by minimizing the KL divergence.
    \item
        Run the HMC algorithm to build a Markov chain of configurations
        following $\tilde{p}(\tilde{\phi}) = e^{-\tilde{S}[\tilde{\phi}]}$,
        \begin{align*}
            \{ \tilde{\phi}_{1},\; \tilde{\phi}_{2},\; \tilde{\phi}_{3},\; \dots ,\;
            \tilde{\phi}_{N} \} \sim e^{-\tilde{S}[\tilde{\phi}]}.
        \end{align*}
    \item
        Apply the inverse transformation $f^{-1}$ to every configuration in the
        Markov chain to undo the variable transformation.  This way we obtain a
        Markov chain of configurations following the target probability
        distribution $p(\phi) = e^{-S[\phi]}$,
        \begin{align*}
            \{ f^{-1}(\tilde{\phi}_{1}),\; f^{-1}(\tilde{\phi}_{2}),\;
            f^{-1}(\tilde{\phi}_{3}),\; \dots ,\; f^{-1}(\tilde{\phi}_{N}) \} =
            \{ \phi_{1},\; \phi_{2},\; \phi_{3},\; \dots ,\; \phi_{N} \} \sim
            e^{-S[\phi]}.
        \end{align*}
\end{enumerate}
The important point is that the acceptance of this algorithm does not depend on
the transformation $f$ that we are doing: it only depends on how well you
integrate the HMC equations of motion. This means that the algorithm will work,
no matter how well one trains the network $f$.

Lüscher proposed this algorithm using the Wilson flow as an approximate
trivializing map \cite{LuscherTrivializingMaps}, but it was not good enough to
improve the scaling of autocorrelation times towards the continuum in a
CP$^{N-1}$ theory with topology \cite{Schaefer2011}. The hope is that
normalizing flows can play a better role as approximate trivializing maps;
indeed, this idea has already been tested \cite{Jin2022,Foreman2022}, but we
want to focus on the scaling of cheap training setups.

\subsection{The model}

As in \cite{Albergo2019,DelDebbio2021}, we worked with a $\phi^{4}$ theory with
a massive scalar field in 2 dimensions,
\begin{align}
    S[\phi]=\sum_{x}^{}\left[-\beta \sum_{\mu =1}^{2}\phi_{x+\mu}\phi _{x}+\phi
_{x}^2+\lambda(\phi_{x}^2-1)^2\right].
\end{align}
Among its features, it has a $\mathbb{Z}_{2}$ symmetry, since the action is
invariant under a sign flip of the scalar field, $\phi \to - \phi$; the
probability density of the model has two modes, which correspond to positive and
negative magnetization $M = \frac{1}{V} \sum_{x}^{} \phi_{x}$; and it has a
non-trivial correlation length $\xi$, yielding autocorrelations when building
the Markov chain of configurations.

We will use these autocorrelations to benchmark our new algorithm against HMC,
since the autocorrelation times of HMC are expected to scale as
$\tau_{\text{int}} \propto \xi^{2}$.  Also, this model does not have topology,
as opposed to QCD, so we will not suffer from topology freezing.

\subsection{Keeping training costs low}

Our intention is to keep training costs as low as possible so that we can say
that the total computational cost is essentially given by the cost of producing
the Markov chain of configurations,
\begin{align}
    \text{total cost} \approx \text{configuration production cost}
\end{align}
For this to happen it is helpful to consider the information we know about the
model:
\begin{itemize}
    \item
        The action has translational symmetry, so there should be no difference
        between the transformation of the field at $x$ and the transformation at
        any other point $x'$. This indicates that one should use convolutional
        neural networks (CNN) instead of fully connected networks, so that the
        same transformation kernel is applied at every point.
    \item
        The relevant physics is contained within the correlation length $\xi$ of
        the system, so the transformation of $\phi_{x}$ should not depend on
        $\phi_{x'}$ if $\left| x - x' \right| \gg \xi$.  This indicates that one
        should make the footprint of the network as small as possible according
        to the correlation length $\xi$.
\end{itemize}
With this in mind we studied very simple network architectures with only one
affine coupling layer and no hidden layers. The affine transformation is
defined as
\begin{align}
    \phi_{x} \to e^{s(\phi)}\phi_{x} + t(\phi),
    \label{eq:afftrafo}
\end{align}
where $s(\phi)$ and $t(\phi)$ are CNNs with kernel size $k$, which can be varied
to control the footprint of the transformation. For the simplest case of $k=3$
and a pair of checkerboard-masked affine layers \cite{Albergo2019}, the
transformation would only couple next-to-nearest neighbors and the neural
network would have only 37 parameters\footnote{A CNN with 1 layer has $k^2$
    parameters. The transformation in Eq.\eqref{eq:afftrafo} with $N_{l}$ pairs
    of checkerboard-masked affine layers has $2\times N_{l}$ different CNNs,
    and therefore $2 \times N_{l} \times k^2$ parameters. We also add a global
    rescaling parameter as a final layer of our network, so our models have
$N_{p} = 2 \times N_{l} \times k^2 + 1$ parameters.}.

With such a simple network the training cost is negligible with respect to the
HMC simulation, and one could compare both algorithms just by studying the
scaling of the autocorrelation time $\tau_{\text{int}}$ towards the continuum.

\section{Results}

\subsection{Reduction of $\tau_{\text{int}}$ with minimal network architecture}

\begin{figure}[!t]
    \centering
    \includegraphics[width=0.49\linewidth]{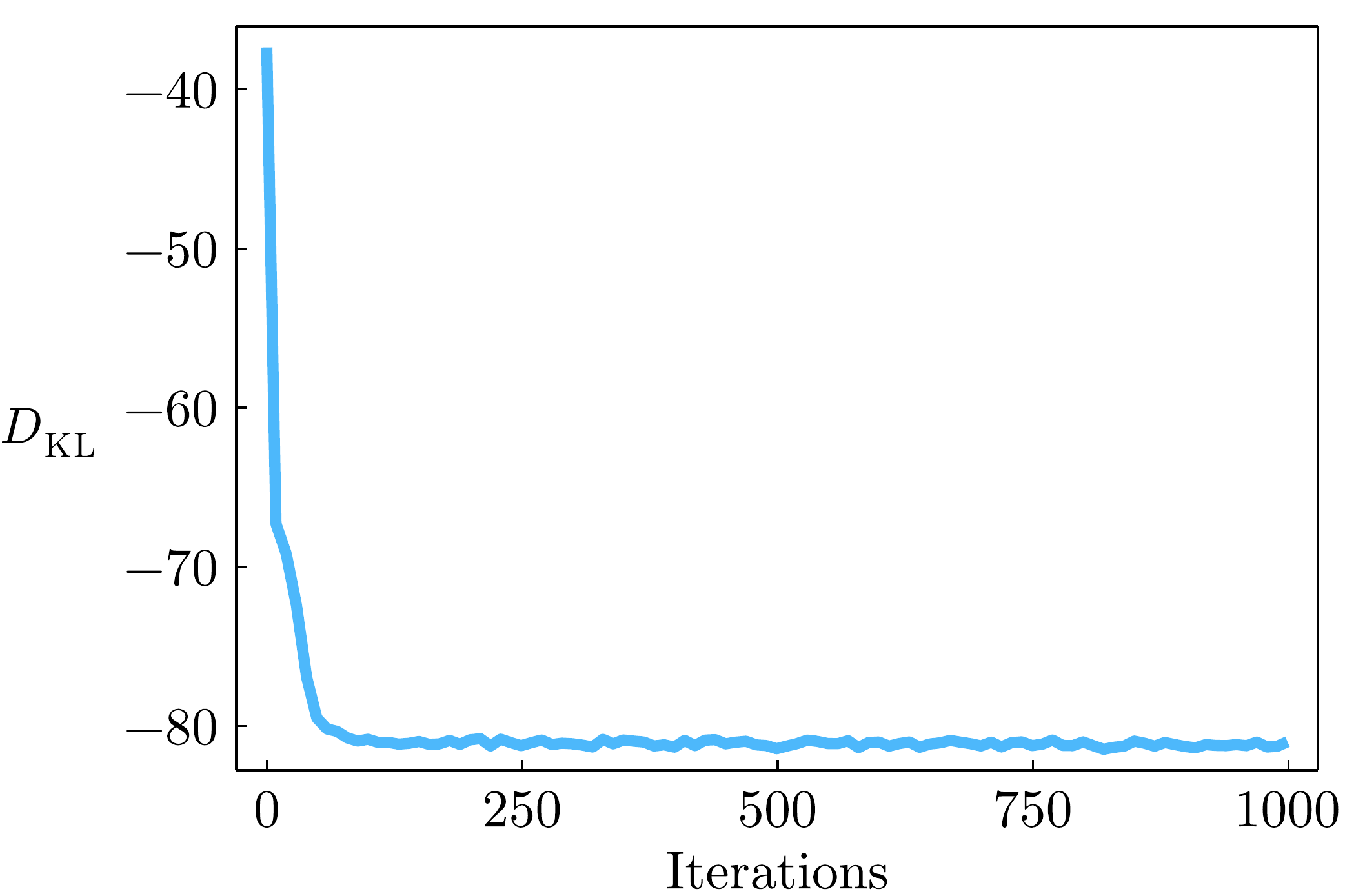}
    \includegraphics[width=0.49\linewidth]{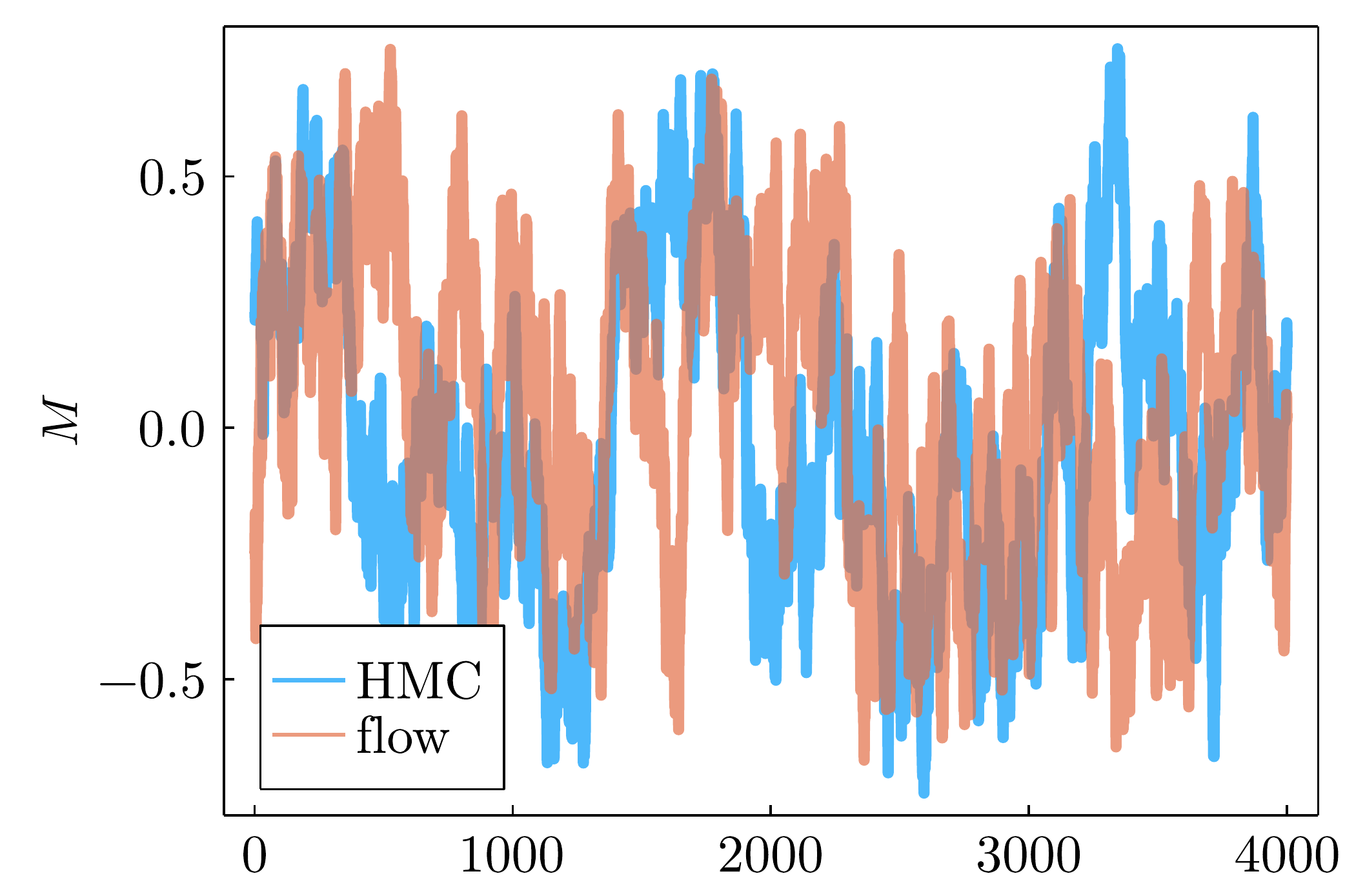}
    \caption{(Left) History of the KL divergence during the training from
        independent Gaussians to a theory with parameters $\beta = 0.641$, $\lambda
    = 0.5$, lattice size $L = 18$ and $\xi = L / 4$. (Right) History of the
magnetization for a simulation with HMC (blue) and trivializing flows (orange).}
    \label{fig:KLandMhist}
\end{figure}

A first thing to check is if a network with only 37 parameters can learn
something.  In Fig.\ref{fig:KLandMhist} (left) we show the evolution of the KL
divergence for the training of a network from independent Gaussians to a theory
with parameters $\beta = 0.641$, $\lambda = 0.5$, lattice size $L = 18$ and $\xi
= L / 4$. We see that $D_{\text{KL}}$ saturates fast, and this is because the
network is very simple: this is the best thing one can do with it, no matter how
much longer is trained.

Having trained the network, one can run a simulation with the two algorithms: one
with plain HMC on the action $S(\phi)$ ---denoted as \emph{HMC} from now on--- and
another with HMC on the transformed action $\tilde{S}(\tilde{\phi})$ ---denoted
as trivializing \emph{flow}. The Monte Carlo history of the magnetization for
both algorithms is shown in Fig.\ref{fig:KLandMhist} (right), and their
respective autocorrelation times are
\begin{align}
    \tau_{\text{flow}} =\; 74.4(3), \quad
    \tau_{\text{HMC}} =\; 100.4(2).
\end{align}
Since the trivializing flows reduced the autocorrelation time of HMC, this
indicates that a simple network can indeed learn something.

\subsection{Infinite volume limit}

\begin{figure}[!t]
    \begin{minipage}{\textwidth}
        \centering
        \begin{minipage}{0.32\textwidth}
            \centering
            \begin{tabular}{rll}
                \(L\) & Acc. at \(L\) & Acc. at \(2L\)\\
                \hline
                3 & 0.3 & 0.2\\
                4 & 0.04 & 0.001\\
                5 & 0.002 & 0.00003\\
                6 & 0.002 & 0.000007\\
                7 & 0.0001 & \(< 10^{-7}\)\\
                8 & 0.0001 & -\\
                9 & 0.00007 & -\\
                10 & 0.00004 & -\\
            \end{tabular}
            \captionof{table}{Metropolis acceptances at lattice sizes $L$ and
            $2L$ of networks trained at lattice size $L$.}
            \label{tab:infiniteV}
        \end{minipage}\hfil
        \begin{minipage}{0.59\textwidth}
            \centering
            \includegraphics[width=\linewidth]{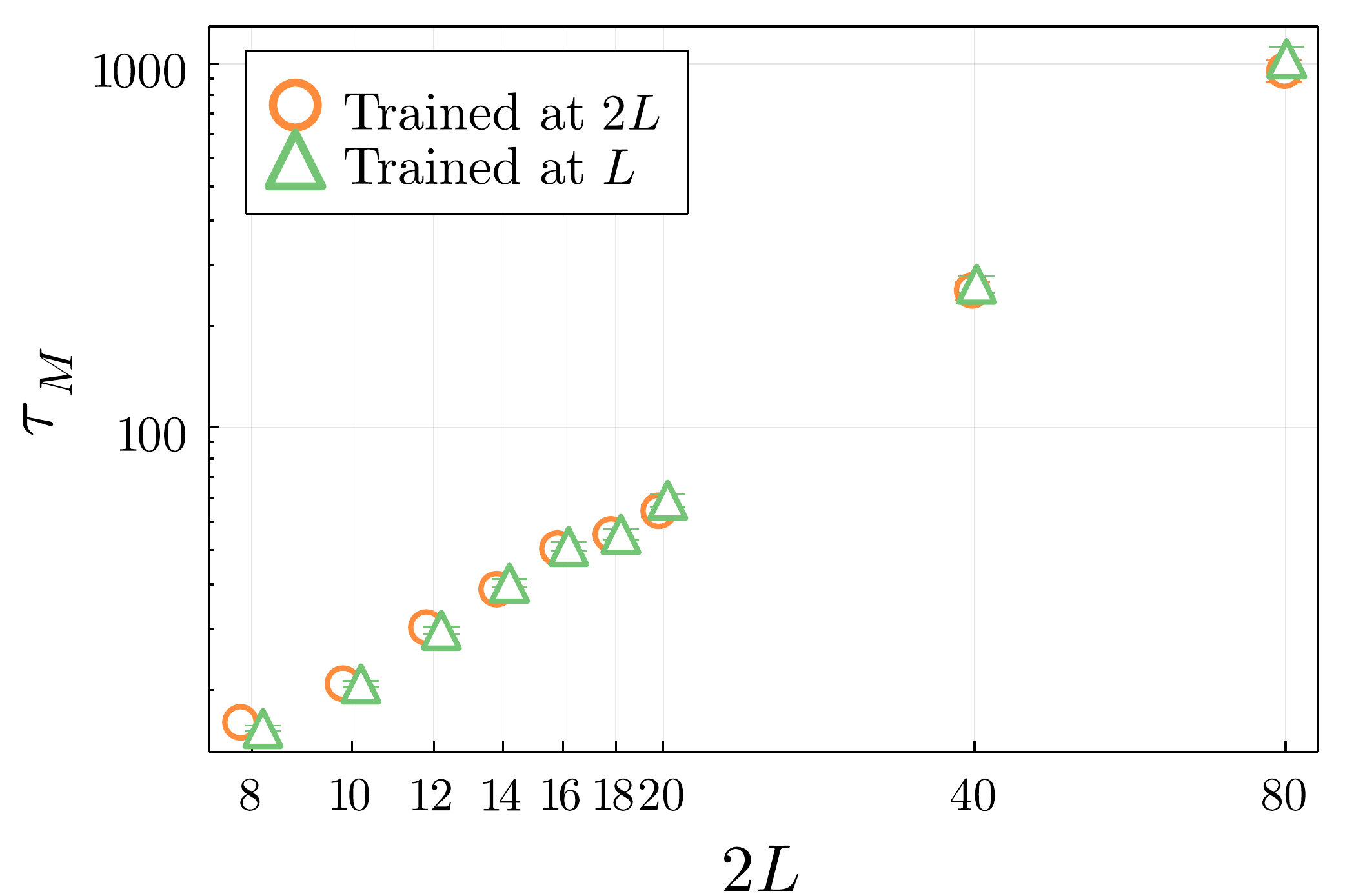}
            \caption{Autocorrelation time of the magnetization at lattice size
                $2L$ using trivializing flows. In circles, the networks used were
                trained at lattice size $2L$; in triangles, they were trained at
            $L$ and used at $2L$.}
            \label{fig:biggerV}
        \end{minipage}\hfill
    \end{minipage}
\end{figure}

Another thing to note is that CNNs allow to take a network trained at a lattice
size $L$ and reuse it for a bigger lattice size $L' > L$. In
Tab.\ref{tab:infiniteV} we trained a different network for every value of $L$
and checked the Metropolis acceptance that they would have using Metropolis
reweighting directly, as in \cite{Albergo2019,DelDebbio2021}. Then, we reused
the network ---without further retraining--- for a lattice size $2L$ and checked
again their Metropolis acceptance.

One can note that the acceptances are low, meaning that the output
distribution $\tilde{p}_{f}$ of our network is not a good approximation of the
target distribution $p$, which was expected because the networks are very
simple; and that reusing the network for a larger lattices decreases the
acceptance a lot, which is also expected because the action is an extensive
quantity and we are just doing reweighting for a bigger volume.

However, none of this matters if the network is used as a trivializing flow. We
can see this in Fig.\ref{fig:biggerV}, where we show the autocorrelation time of
the magnetization versus different lattice sizes; whether you train directly at
a lattice size $2L$ or you train at $L$ and reuse it for $2L$, you get the same
autocorrelation times. This indicates that the training should be done at a
lattice size similar to the correlation length of the system, $L \sim \xi$,
where the physical information is contained, to reduce training costs.

\subsection{Computational cost scaling}

% \begin{center}
% \begin{tabular}{rrrl}
% \(\lambda\) & \(L\) & \(\beta\) & Network acc.\\
% \hline
% 0.5 & 6 & 0.537 & 0.3\\
% 0.5 & 8 & 0.576 & 0.04\\
% 0.5 & 10 & 0.601 & 0.002\\
% 0.5 & 12 & 0.616 & 0.002\\
% 0.5 & 14 & 0.626 & 0.0001\\
% 0.5 & 16 & 0.634 & 0.0001\\
% 0.5 & 18 & 0.641 & 0.00007\\
% 0.5 & 20 & 0.645 & 0.00004\\
% 0.5 & 40 & 0.667 & -\\
% 0.5 & 80 & 0.677 & -\\
% \end{tabular}
% \end{center}

\subsubsection{The setup}

Finally, we study how the computational cost of the machine-learned trivializing
flow scales towards the continuum, comparing it with the costs of HMC. The
theory parameters used are in the table below, where $\beta$ was tuned to fix the
physical size of the lattice to $\xi = L / 4$.

\begin{center}
\begin{tabular}{lrrrrrrrrrr}
    \hline
    \hline
\(L\) & 6 & 8 & 10 & 12 & 14 & 16 & 18 & 20 & 40 & 80\\
\(\beta\) & 0.537 & 0.576 & 0.601 & 0.616 & 0.626 & 0.634 & 0.641 & 0.645 & 0.667 & 0.677\\
\(\lambda\) & 0.5 & 0.5 & 0.5 & 0.5 & 0.5 & 0.5 & 0.5 & 0.5 & 0.5 & 0.5\\
\hline
\hline
\end{tabular}
\end{center}
A few things are worth noting:
\begin{itemize}
    \item
        For each of the columns we train a different network from independent
        Gaussians to the target theory.
    \item
        We always used simple network architectures with one coupling layer and
        no hidden layers, and we trained them until saturation of the KL
        divergence. This saturation happens very fast and the training costs are
        negligible with respect to the cost of the production of configurations.
        Therefore we only need to compare the scaling of the autocorrelation
        times of both algorithms.
    \item
        The integration step for the molecular dynamics evolution is tuned so
        that the Metropolis--Hastings acceptance of HMC and the trivializing
        flows is approximately 90\%.
\end{itemize}

\begin{figure}[!t]
    \centering
    \includegraphics[width=0.49\linewidth]{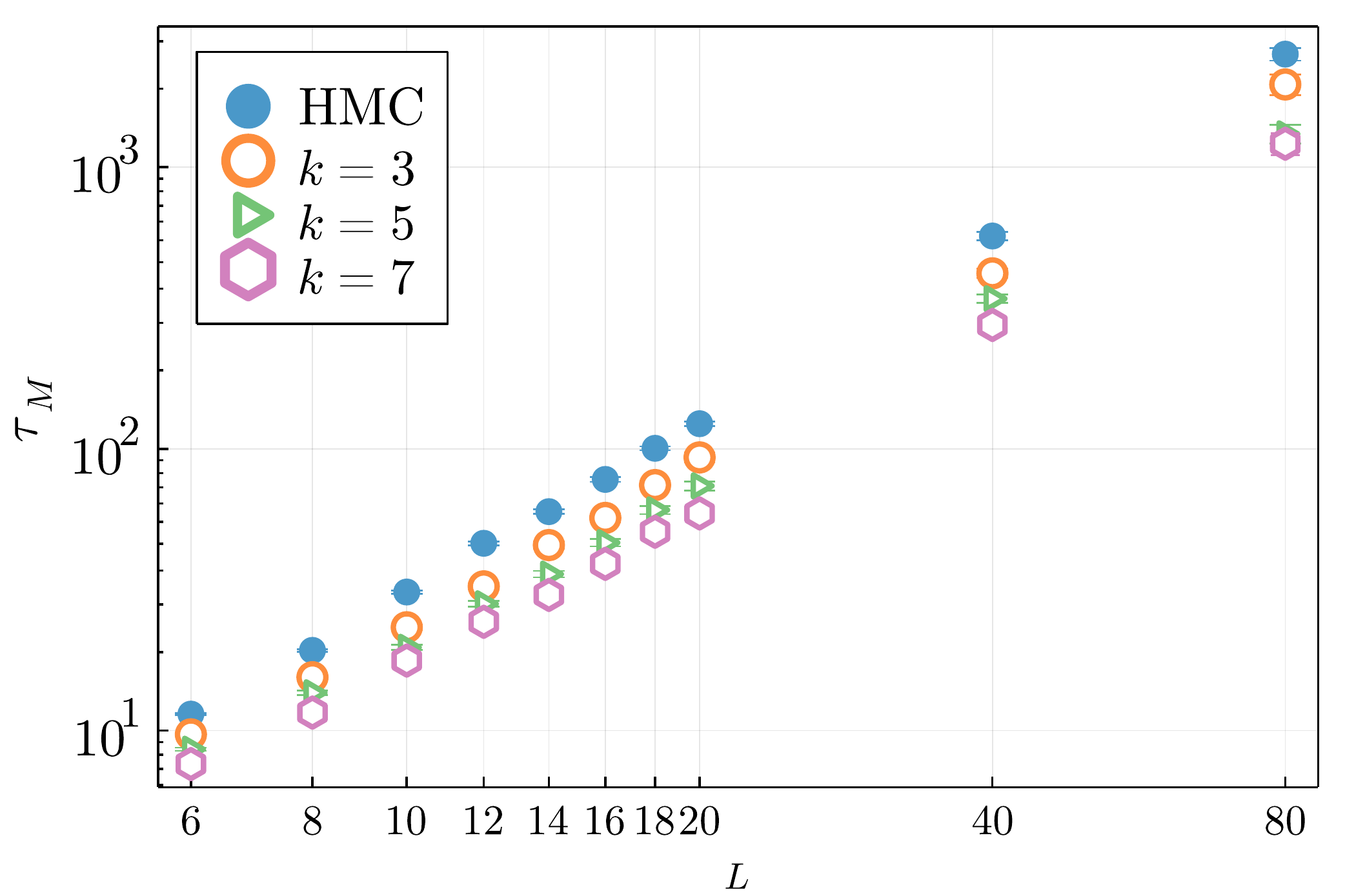}
    \includegraphics[width=0.49\linewidth]{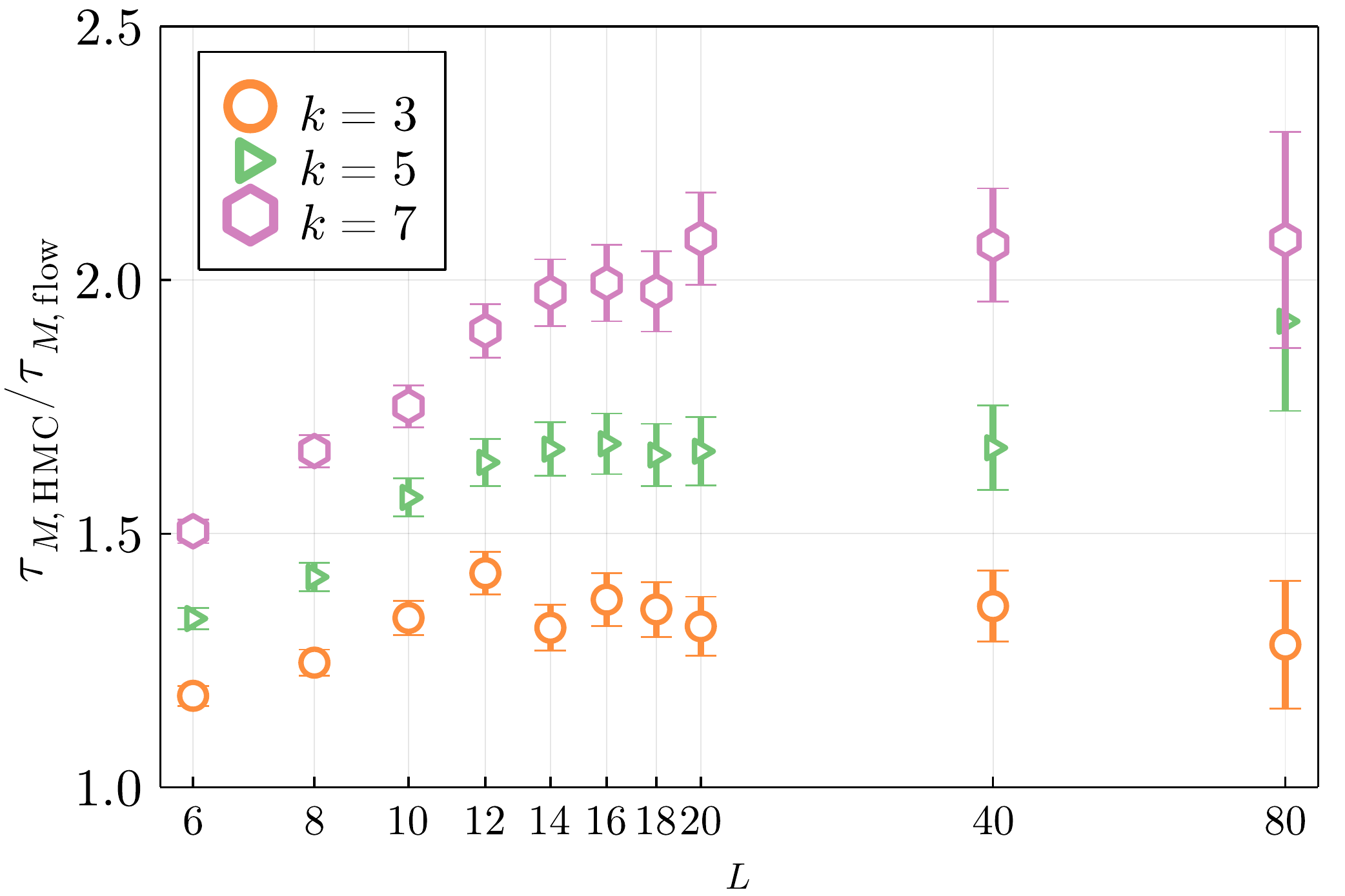}
    \caption{(Left) Scaling of the autocorrelation time of the magnetization
        towards the continuum for HMC (filled, blue circles) and trivializing
        flows with kernel sizes $k=3,5,7$ (open circles, triangles and
        hexagons).  (Right) Scaling of the ratio of autocorrelation times of the
    magnetization of HMC with respect to trivializing flows.} 
    \label{fig:Mfixedarch}
\end{figure}

\subsubsection{Scaling with fixed architecture}

In Fig.\ref{fig:Mfixedarch} (left) we compare the autocorrelation times of the
magnetization from simulations with both algorithms: in blue circles we have the
autocorrelations of HMC, and the rest of points are from trivializing flow
simulations with kernel sizes $k=3, 5, 7$.

First, we see that all autocorrelations from trivializing flows are better than
the ones of HMC. Also, for a fixed lattice size $L$, increasing the kernel size
of the network decreases the autocorrelation, but it does not change the scaling
of HMC. 

The latter can be seen better if we plot the ratio $\tau_{\text{HMC}} /
\tau_{\text{flow}}$, as we do in Fig.\ref{fig:Mfixedarch} (right): as we go to
the continuum the ratio tends to a constant, indicating that the scaling of both
algorithms is the same for a fixed network architecture.

\subsubsection{Scaling increasing the kernel size}

\begin{figure}[!t]
	\centering
    \includegraphics[width=0.49\linewidth]{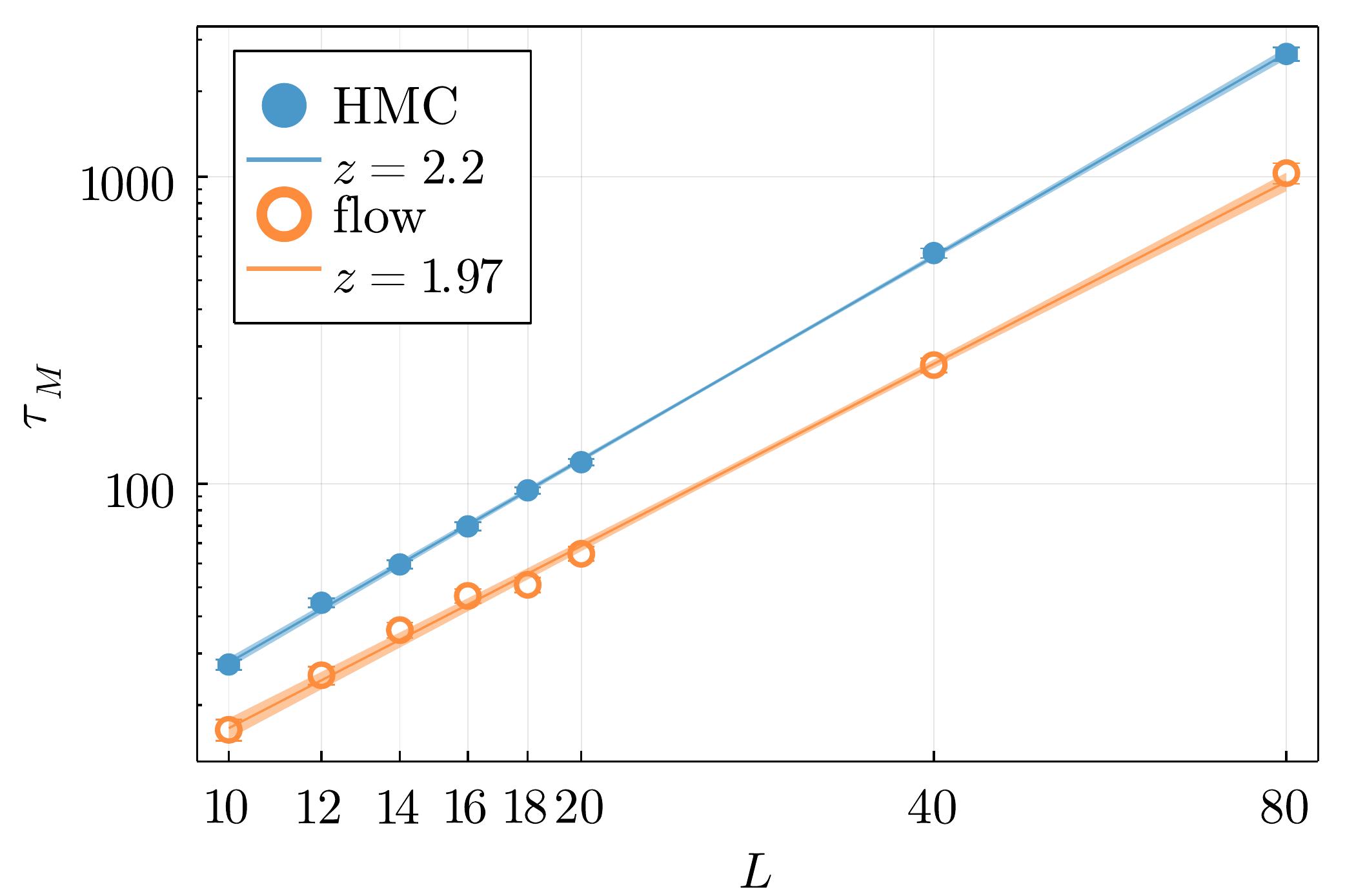}
    \includegraphics[width=0.49\linewidth]{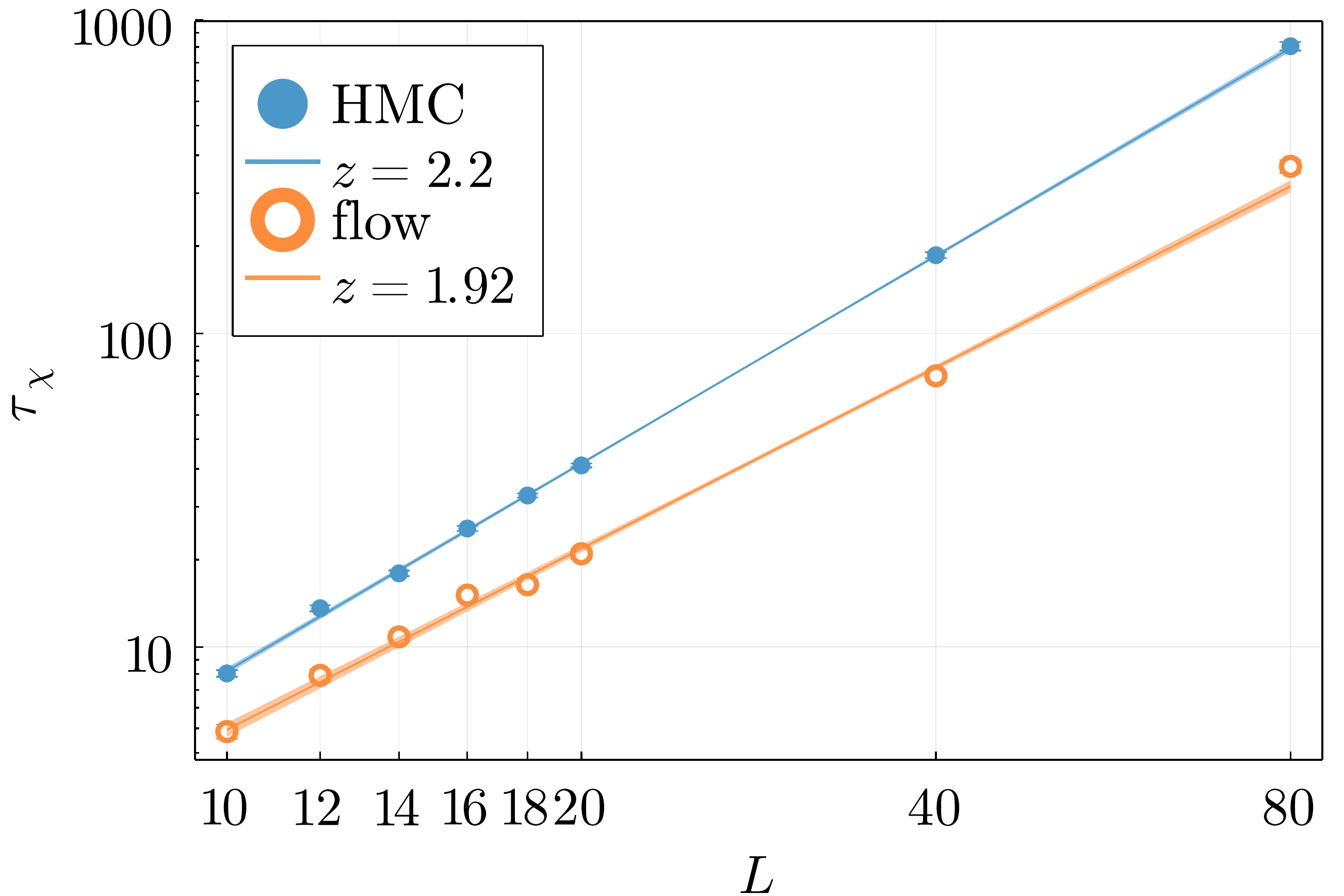}
    \caption{(Left) Scaling of the autocorrelation time of the magnetization
        with $k\sim\xi$. (Right) Scaling of the autocorrelation time of the
        flowed one-point susceptibility with $k\sim\xi$. The fits correspond to
    the assumption $\tau_{\text{int}} \propto \xi^{z}$.}
    \label{fig:tauM-scalingk}
\end{figure}

The correlation length $\xi$ increases in lattice units when taking the
continuum limit, so one should also scale the kernel size of our transformations
so that the transformation acts on the same physical region, $k \sim \xi $. When
one chooses the optimum kernel size $k$ of the CNNs for each lattice size, one
gets the results displayed in Fig.\ref{fig:tauM-scalingk} (left), where now the
gap between HMC and the trivializing flow algorithm seems to increase towards
the continuum. 

Assuming that the autocorrelation length scales as $\tau \propto \xi^{z}$ we can
fit our results and compare the exponent of the two algorithms, obtaining 
\begin{align}
    z_{M, \text{HMC}} =\; 2.20(4), \quad z_{M, \text{flow}} =\; 1.97(7).
\end{align}
Therefore, scaling the kernel size of the CNNs leads to a slight improvement in
the autocorrelation scaling of the magnetization $M = 1 / V \sum_{x}^{}
\phi_{x}$, which is a local operator. In Fig.\ref{fig:tauM-scalingk} (right) we
plot instead the one-point susceptibility $\chi_{t} = \frac{1}{V} \sum_{x}^{}
\phi_{t,x}^{2}$ measured on configurations flowed up to flow time $t$ with a
smearing radius $R \sim \xi$. Assuming the same scaling we again find a slight
improvement with respect to HMC:
\begin{align}
    z_{\chi_{t}, \text{HMC}} =\; 2.20(2), \quad
    z_{\chi_t, \text{flow}} =\; 1.92(4).
\end{align}

\section{Conclusions and outlook}

We have shown that, even working with very simple architectures, using neural
networks as trivializing flows can improve the autocorrelation times of HMC,
although the scaling is the same as HMC for a fixed network architecture.

Also, the networks trained at a small lattice size can be reused for larger
volumes without further training. Focusing on topology freezing, this could be
useful in QCD: one could train at the size of $\Lambda_{\text{QCD}}$ for large
values of the quark masses and then reuse the network for larger volumes and
small masses, thus reducing training costs.

Finally, scaling the kernel size of the networks slightly improves the scaling
of autocorrelations. An interesting question would then be if machine-learned
trivializing flows could improve a much worse kind of continuum scaling, such as
the one related to topological freezing in theories with topology.

\section*{Acknowledgments}
\addcontentsline{toc}{section}{Acknowledgements}

We acknowledge support from the Generalitat Valenciana grant PROMETEO/2019/083,
the European project H2020-MSCA-ITN-2019//860881-HIDDeN, and the national
project PID2020-113644GB-I00. AR acknowledges financial support from Generalitat
Valenciana through the plan GenT program (CIDEGENT/2019/040). DA acknowledges
support from the Generalitat Valenciana grant ACIF/2020/011. JMR is supported by
STFC grant ST/T506060/1.

This work has been performed under the Project HPC-EUROPA3
(INFRAIA-2016-1-730897), with the support of the EC Research Innovation Action
under the H2020 Programme; in particular, we gratefully acknowledge the support
of the computer resources and technical support provided by EPCC. This work used
the ARCHER2 UK National Supercomputing Service
(\url{https://www.archer2.ac.uk}).  We also acknowledge the computational
resources provided by Finis Terrae II (CESGA), Lluis Vives (UV), Tirant III
(UV). The authors also gratefully acknowledge the computer resources at
Artemisa, funded by the European Union ERDF and Comunitat Valenciana, as well as
the technical support provided by the Instituto de Física Corpuscular, IFIC
(CSIC-UV).

% \begin{thebibliography}{99}
% \bibitem{...}
% ....

% \end{thebibliography}

\bibliography{references}
\addcontentsline{toc}{section}{References}
\bibliographystyle{unsrt}

\end{document}